\documentclass[twocolumn,preprintnumbers,amsmath,amssymb,prb]{revtex4}

\usepackage{graphicx}
\usepackage{amssymb,amsmath}
\usepackage{dcolumn}
\usepackage{bm}
\usepackage{color}
\usepackage{enumerate}
\usepackage{epstopdf}

\begin{document}

\title{Light-induced bound electron states in two-dimensional systems: \\Contribution to electron transport}
\author{O. V. Kibis}\email{Oleg.Kibis(с)nstu.ru}
\author{M. V. Boev}
\author{V. M. Kovalev}

\affiliation{Department of Applied and Theoretical Physics,
Novosibirsk State Technical University, Karl Marx Avenue 20,
Novosibirsk 630073, Russia}

\begin{abstract}

In two-dimensional (2D) electron systems, an off-resonant
high-frequency circularly polarized electromagnetic field can
induce the quasi-stationary bound electron states of repulsive
scatterers. As a consequence, the resonant scattering of
conduction electrons through the quasi-stationary states and the
capture of conduction electrons by the states appear. The present
theory describes the transport properties of 2D electron gas
irradiated by a circularly polarized light, which are modified by
these processes. Particularly, it is demonstrated that irradiation
of 2D electron systems by the off-resonant field results in the
quantum correction to conductivity of resonant kind.

\end{abstract}

\maketitle

\section{Introduction}
The control of physical properties of various quantum systems by
an off-resonant high-frequency electromagnetic field became the
important and established research area which resulted in many
fundamental effects (see, e.g.,
Refs.~\onlinecite{Goldman_2014,Bukov_2015,Casas_2011,Eckardt_2015,Rahav_2003,Holthaus_2015,Basov_2017,Vogl_2019,
Vogl_2020}). Since the frequency of the field lies far from
characteristic resonant frequencies of an irradiated system
(off-resonant field), the field can not be absorbed by electrons
and only interacts non-resonantly with them (``dresses'' them). As
a result, the behavior of dressed electrons varies as a function
of the dressing field. To clarify the field-induced features of
electronic properties of low-dimensional systems, many works
dedicated to the electromagnetic dressing of various
nanostructures --- including quantum
rings~\cite{Kibis_2011,Koshelev_2015,Kozin_2018,Kozin_2018_1},
quantum
wells~\cite{Yin_2011,Morina_2015,Pervishko_2015,Dini_2016,Avetissian_2016},
topological
insulators~\cite{Lindner_2011,Rechtsman_2013,Wang_2013,Torres_2014,Usaj_2014,Dehghani_2015,Calvo_2015,Morimoto_2016,Mikami_2016,Kyriienko_2019},
graphene and related 2D
materials~\cite{Oka_2009,Kibis_2010,Syzranov_2013,Glazov_2014,Perez_2014,Sentef_2015,Sie_2015,Kibis_2016,Kibis_2017,Iorsh_2017,Sato_2019,Seifert_2019,Iurov_2019,Iurov_2020,Cavalleri_2020},
etc --- were published.

Among variety of phenomena in periodically driven systems, the
effect of dynamical stabilization should be noted especially.
Generally, the dynamical stabilization is the fundamental physical
effect which consists in the stabilization of initially unstable
systems by oscillating external actions (see, e.g.,
Ref.~\onlinecite{Bukov_2015}). Phenomenologically, this effect was
known for a long time to any circus juggler who hold various
objects in balance by vibrational movements of its hands. First
correct description of the effect in the frame of classical
Hamilton mechanics was done by P. L. Kapitza who suggested the
simple mechanical model based on a pendulum~\cite{Kapitza_1951}.
The Kapitza pendulum is the system consisting of a point mass
attached to a light inextensible rod connected to a vibrating
suspension. In the case of fixed point of suspension, this model
describes the conventional mathematical pendulum for which there
are two points of equilibrium (the lower and the upper). The
equilibrium of the pendulum in the upper point is unstable and any
infinitesimal perturbation leads to loss of the equilibrium.
However, in contravention of intuition, the upper (vertical)
position of the pendulum can be steady in the case of fast
oscillating suspension. Thus, the Kapitza pendulum acquires the
quasi-stable equilibrium (the local minimum of its potential
energy) at the upper point which corresponds to the maximum of its
potential energy in the absence of oscillations. This effect is of
universal nature and occurs in many different areas of physics.
For example, an oscillating laser field can lead to stabilization
of charged ionic systems whose components repulse each other
according to the Coulomb law and, therefore, cannot exist as a
whole without the field~\cite{Duijn_1996}. Despite the long
prehistory of the effect of dynamical stabilization, its possible
manifestations in nanostructures still wait for detailed study. To
fill partially this gap in the theory, we analyzed recently the
behavior of various repulsive potentials in nanostructures driven
by an oscillating field in context of the dynamical
stabilization~\cite{Kibis_2019}. Particularly, it was found that a
circularly polarized electromagnetic field can induce the local
minima of potential energy in the core of 2D repulsive potentials.
As a consequence, the quasi-stationary electron states confined
near the local minima appear. The present article is dedicated to
theoretical analysis of the electron transport in 2D systems
modified by the light-induced quasi-stationary electron states
bound at repulsive scatterers.

The article is organized as follows. In Sec.~II, we develop the
theory of the light-induced quasi-stationary electron states bound
at short-range scatterers modeled by the repulsive
delta-potential. In Sec.~III, the Boltzmann kinetic equation
taking into account the scattering and capture of conduction
electrons by the quasi-stationary bound states is solved. In
Sec.~IV, analysis of the conductivity of the irradiated 2D
electron system is performed. The last two sections of the article
contain conclusion and acknowledgments.

\section{Model}
Let us consider a 2D electron system in the $(x,y)$ plane
irradiated by a circularly polarized electromagnetic wave
propagating along the $z$ axis with the vector potential
$\mathbf{A}(t)=(A_x,A_y)=[cE_0/\omega](\sin\omega t,\,\cos\omega
t)$, where $E_{0}$ is the electric field amplitude of the
wave, and $\omega$ is the wave frequency which lies far from
characteristic resonant frequencies of the electron system (the
off-resonant dressing field) [see Fig.~1a]. The physical
properties of the dressed 2D electron system in the presence of a
scatterer with the repulsive potential $U(\mathbf{r})$ are
described by the Hamiltonian
\begin{equation}\label{HA}
\hat{\cal
H}=\frac{[\hat{\mathbf{p}}-e\mathbf{A}(t)/c]^2}{2m_e}+U(\mathbf{r}),
\end{equation}
where $\hat{\mathbf{p}}=(\hat{p}_x,\hat{p}_y)$ is the plane
momentum operator, $m_e$ is the effective electron mass, $e$ is
the electron charge,  and $\mathbf{r}=(x,y)$ is the plane radius
vector of an electron. Applying the Kramers-Henneberger unitary
transformation~\cite{Kramers_52,Henneberger_68},
\begin{equation}\label{KHU}
\hat{\cal
U}(t)=\exp\left\{\frac{i}{\hbar}\int^{\,t}_{-\infty}\left[
\frac{e}{m_ec}\mathbf{A}(t^\prime)\hat{\mathbf{p}}-\frac{e^2}{2m_ec^2}A^2(t^\prime)
\right]dt^\prime\right\},
\end{equation}
the transformed Hamiltonian (\ref{HA}) reads
\begin{eqnarray}\label{HT}
\hat{\cal H}^{\prime}&=&\hat{\cal U}^\dagger(t)\hat{\cal
H}\hat{\cal U}(t) - i\hbar\hat{\cal U}^\dagger(t)\partial_t
\hat{\cal U}(t)\nonumber\\
&=&\frac{\hat{\mathbf{p}}^2}{2m_e}+U(\mathbf{r}-\mathbf{r}_0(t)),
\end{eqnarray}
where the radius vector $\mathbf{r}_0(t)=(-r_0\cos\omega
t,\,r_0\sin\omega t)$ describes the classical circular trajectory
of electron movement in the circularly polarized field, and
\begin{equation}\label{r0}
r_0=\frac{|e|E_0}{m_e\omega^2}
\end{equation}
is the radius of the trajectory~\cite{Landau_2}.
\begin{figure}[!ht]
\includegraphics[width=0.7\columnwidth]{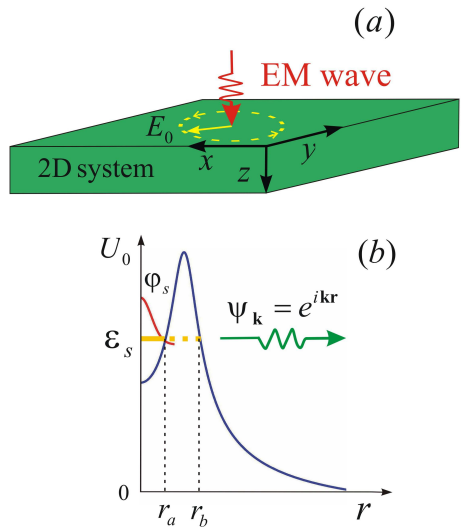}
\caption{Sketch of the system under consideration: (a) 2D electron
system irradiated by a circularly polarized electromagnetic wave
with the electric field amplitude $E_0$; (b) The 2D scattering
repulsive potential dressed by a circularly polarized field,
$U_0(r)$, which has the local minimum at $r=0$ and contains the
bound electron state with the wave function $\varphi_s$ (the
Gaussian-like red line) and the energy $\varepsilon_s$ (the
horizontal yellow line). The bound state is quasi-stationary and
can decay due to the tunneling through the potential barrier
between the two turning points, $r_a$ and $r_b$. As a result of
the tunneling, the emission of free electron with the wave
function $\psi_\mathbf{k}=e^{i\mathbf{kr}}$ (the green wave arrow)
from the bound state takes place.}\label{Fig.1}
\end{figure}
Physically, the unitary transformation (\ref{KHU}) corresponds to
the transition from the laboratory reference frame to the rotating
reference frame,
\begin{equation}\label{rf}
\mathbf{r}\rightarrow\mathbf{r}-\mathbf{r}_0(t).
\end{equation}
Expanding the oscillating potential in Eq.~(\ref{HT}) into a
Fourier series,
\begin{equation}\label{Uosc}
U(\mathbf{r}-\mathbf{r}_0(t))=U_0(\mathbf{r})+\left[\sum_{n=1}^\infty
U_n(\mathbf{r})e^{in\omega t}+\mathrm{c.\,c.}\right],
\end{equation}
the transformed Hamiltonian (\ref{HT}) can be rewritten as
\begin{equation}\label{H3}
\hat{\cal
H}^\prime=\frac{\hat{\mathbf{p}}^2}{2m_e}+U_0(\mathbf{r})+\left[\sum_{n=1}^\infty
U_n(\mathbf{r})e^{in\omega t}+\mathrm{c.\,c.}\right],
\end{equation}
where
\begin{equation}\label{U0}
U_0(\mathbf{r})=\frac{1}{2\pi}\int_{-\pi}^{\pi}U\big(\mathbf{r}-\mathbf{r}_0(t)\big)\,d(\omega
t)
\end{equation}
is the stationary part of the potential, which is responsible for
the smooth motion of 2D electrons, and $U_n(\mathbf{r})$ are the
Fourier coefficients of the oscillating potential. The stationary
potential (\ref{U0}) should be treated as a repulsive potential
dressed by an oscillating field (dressed potential). The specific
feature of the dressed potential (\ref{U0}) is the local minimum
existing near $r=0$ if the field is strong
enough~\cite{Kibis_2019} (see Fig.~1b). As a consequence of the
local minimum, the domain of attraction takes place in the core of
the repulsive potential. This attraction results in confinement of
an electron at the repulsive potential and the bound electron
state with the energy $\varepsilon_s$ and the localized wave
function $\varphi_s$ appears. In the following, we will assume
that the field frequency, $\omega$, satisfies the two conditions:
Firstly,
\begin{equation}\label{taue}
\omega\tau_e\gg1,
\end{equation}
where $\tau_e$ is the mean free time of conduction electrons;
secondly, the field frequency lies far from resonant frequencies
of the bound state. Under the first condition, scattering
processes cannot destroy the bound state, whereas the second
condition allows to neglect the effect of the oscillating terms of
the Hamiltonian (\ref{H3}) on the bound state. As a consequence,
the electron dynamics of an electron confined at a scatterer can
be described solely by the stationary part of the
Hamiltonian,~\cite{Kibis_2019}
\begin{equation}\label{H0}
\hat{\cal H}_0=\frac{\hat{\mathbf{p}}^2}{2m_e}+U_0(\mathbf{r}).
\end{equation}

The field-induced local minimum of the dressed potential
$U_0(\mathbf{r})$ forms the quantum well which separates the bound
electron state inside the well, $\varphi_s$, from the states of
free conduction electrons outside the well,
$\psi_\mathbf{k}=e^{i\mathbf{k}\mathbf{r}}$, by the potential
barrier between the turning points, $r_a$ and $r_b$ (see Fig.~1b).
Physically, the electron tunneling through the barrier,
$\varphi_s\rightarrow\psi_\mathbf{k}$, results in finite lifetime
of the bound state and the broadening of its energy. To find the
bound state energy $\varepsilon$ and the energy broadening
$\Gamma$, we have to solve the Schr\"odinger equation, $\hat{\cal
H}_0\psi=\varepsilon\psi$, with the Hamiltonian (\ref{H0}). Let us
model the repulsive potential $U(\mathbf{r})$ with the delta
function,
\begin{equation}\label{UU}
U(\mathbf{r})=u_0\delta(\mathbf{r}),
\end{equation}
where $u_0>0$ is the strength of the repulsive potential.
Substituting the potential (\ref{UU}) into Eq.~(\ref{U0}), we
arrive at the dressed delta-potential,
\begin{equation}\label{U00}
U_0(\mathbf{r})=\frac{u_0\,\delta({r}-{r}_0)}{2\pi r_0}.
\end{equation}
Thus, the circularly polarized field turns the repulsive delta
potential (\ref{UU}) into the delta potential barrier of ring
shape (\ref{U00}). As a consequence, quasi-stationary electron
states confined inside the area fenced by the ring-shape barrier
($0<r<r_0$) can exist. Substituting the dressed potential
(\ref{U00}) into Eq.~(\ref{H0}), we arrive at the Hamiltonian
\begin{equation}\label{H1}
\hat{\cal
H}_0=-\frac{\hbar^2}{2m_er}\left[\frac{\partial}{\partial
r}\left(r\frac{\partial}{\partial
r}\right)+\frac{1}{r}\frac{\partial^2}{\partial\varphi^2}\right]+\frac{u_0\,\delta({r}-{r}_0)}{2\pi
r_0},
\end{equation}
where $\varphi$ is the azimuth angle in the 2D plane. The
eigenfunction of the Hamiltonian, $\psi$, which corresponds to an
electron confined in the area $0<r<r_0$, must be finite at $r=0$
and satisfy the condition $\psi|_{r\rightarrow\infty}\propto
e^{ikr}$, where $k=\sqrt{\,2m_e\varepsilon}/\hbar$ is the electron
wave vector. Solving the Schr\"odinger problem with the
Hamiltonian (\ref{H1}) under these conditions and assuming
$\alpha=2\hbar^2/m_eu_0\ll1$ (see Appendix A), we arrive at the
sought energy spectrum of an electron confined inside the
ring-shape delta-barrier,
$\varepsilon=\varepsilon_{nm}-i\Gamma_{nm}/2$, and the
corresponding wave functions, $\psi_{nm}$, where the energy of
quasi-discrete electron level is
\begin{equation}\label{eeeE}
\varepsilon_{nm}=\frac{\hbar^2\bar{z}^2_{nm}}{2m_er^2_0}+{\cal
O}\left(\alpha\right),
\end{equation}
the broadening of the energy level is
\begin{equation}\label{G2Ee}
{\Gamma}_{nm}=\frac{4\varepsilon_{nm}\alpha^2}{N^3_m(\bar{z}_{nm})[J_{m+1}(\bar{z}_{nm})-J_{m-1}(\bar{z}_{nm})]}+{\cal
O}\left(\alpha^3\right),
\end{equation}
the wave functions are
\begin{eqnarray}\label{Flm}
\psi_{nm}&=&\frac{e^{im\varphi}}{\sqrt{\pi}r_0J_{m+1}(\bar{z}_{nm})}\left\{
\begin{array}{rl}
J_m\left(\frac{\bar{z}_{nm}r}{r_0}\right), &0<r\le r_0\\
0, &r\ge r_0
\end{array}\right.\nonumber\\
&+&{\cal O}\left(\alpha\right),
\end{eqnarray}
$J_m(\xi)$ is the $m$th Bessel function of the first kind,
$N_m(\xi)$ is the  $m$th Bessel function of the second kind (the
Neumann function), $\bar{z}_{nm}$ is the $n$th zero of the $m$th
Bessel function of the first kind (i.e. $J_m(\bar{z}_{nm})=0$),
and $n=1,2,3,...$ is the principal quantum number. As expected, in
the limiting case of $u_0\rightarrow\infty$, the energy broadening
(\ref{G2Ee}) is zero and the quasi-stationary electron state
(\ref{eeeE}) turns into the stationary one.

Generally, the discussed bound state exists if its energy
broadening is small as compared with the characteristic distance
between the neighbor energy levels. Therefore, we will restrict
the following consideration by the ground bound state,
$\varphi_s(r)=\psi_{10}(r)$, which has the minimal broadening
$\Gamma_s=\Gamma_{10}$. It follows from
Eqs.~(\ref{eeeE})--(\ref{G2Ee}) that the energy of the ground bound
state and its broadening read
\begin{equation}\label{ee0E}
\varepsilon_{s}=\frac{\hbar^2\bar{z}^2_{10}}{2m_er^2_0},
\end{equation}
\begin{equation}\label{G2E}
{\Gamma}_{s}=\frac{2\,\varepsilon_{s}\,\alpha^2}{N^3_0(\bar{z}_{10})J_1(\bar{z}_{10})},
\end{equation}
where $\bar{z}_{10}\approx2.4$ is the first zero of the zeroth
Bessel function of the first kind. Correspondingly, the
applicability condition of Eqs.~(\ref{ee0E})--(\ref{G2E}) can be
written as
\begin{equation}\label{Ge}
\frac{\Gamma_s}{\varepsilon_s}\ll1.
\end{equation}
The condition (\ref{Ge}) can be satisfied if the scattering
potential (\ref{UU}) is strong enough, i.e.
$\alpha=2\hbar^2/m_eu_0\ll1$. It should be noted that this
condition does not depend on the strength of the alternating field
because of the delta-function singularity of the model repulsive
potential (\ref{UU}). If the repulsive potential $U(\mathbf{r})$
is described by any smooth function, the quasi-stationary state
--- or, what is the same, the field-induced local minimum of
potential energy in the core of the repulsive potential --- takes
place if the alternating field is strong enough~\cite{Kibis_2019}.

\section{Kinetic equation}

The field-induced modification of a repulsive potential discussed
above results in both capture of conduction electrons at the bound
state and to scattering them through the quasi-discrete energy
level $\varepsilon_s$. Let us analyze the effect of these
processes on electron transport under the stationary field
$\mathbf{E}$ applied to the 2D system of the area $S$. In the
laboratory reference frame, the conventional Boltzmann kinetic
equation (see, e.g., Ref.~\onlinecite{Ziman_book}) can be written
as
\begin{equation}
\frac{\partial f_{\mathbf{k}}}{\partial
t}\Bigg]_{\mathrm{field}}+\frac{\partial f_{\mathbf{k}}}{\partial
t}\Bigg]_{\mathrm{scatter}}=0,\label{Be1}
\end{equation}
where $f_{\mathbf{k}}$ is the distribution function of conduction
electrons with the wave vector $\mathbf{k}$,
\begin{equation}\label{fieldc}
\frac{\partial f_{\mathbf{k}}}{\partial
t}\Bigg]_{\mathrm{field}}=-\frac{\partial
f_{\mathbf{k}}}{\partial\mathbf{k}}\,\frac{e\mathbf{E}}{\hbar}
\end{equation}
is the field term describing the transport of conduction electrons
under the weak stationary electric field $\mathbf{E}$,
\begin{equation}\label{sct}
\frac{\partial f_{\mathbf{k}}}{\partial
t}\Bigg]_{\mathrm{scatter}}=\sum_{\mathbf{k^\prime}}[f_{\mathbf{k}^\prime}
(1-f_{\mathbf{k}})-f_{\mathbf{k}}
(1-f_{\mathbf{k}^\prime})]w_{\mathbf{k}\mathbf{k}^\prime}
\end{equation}
is the term describing the scattering of conduction electrons, and
$w_{\mathbf{k}\mathbf{k}^\prime}$ is the probability of electron
scattering between the states $\mathbf{k}^\prime$ and $\mathbf{k}$
per unit time. In the following, we will assume that the
light-induced bound states of scatterers can capture only one
electron per scatterer because of the Coulomb repulsion between
electrons. Then the conservation law for the total number of
electrons in the 2D system reads
\begin{equation}
N_e+N_sf_s=N_0,\label{Be2}
\end{equation}
where $N_0$ is the total number of conduction electrons in the 2D
system in the absence of irradiation,
\begin{equation}\label{Be3}
N_e=2\sum_{\mathbf{k}}f_{\mathbf{k}}
\end{equation}
is the total number of conduction electrons in the irradiated 2D
system, and $N_sf_s$  is the total number of electrons captured by
scatterers. Correspondingly, $N_s$ is the total number of
scatterers in the 2D system, and $f_s$ is the distribution
function of conduction electrons captured by the scatterers. In
Eq.~(\ref{Be3}) and what follows, the sum symbol $\sum_\mathbf{k}$
denotes the summation over all electron states with different wave
vectors $\mathbf{k}$, excluding the summation over the spin
freedom degrees [see Eq.~(\ref{sumk}) in Appendix B].

The distribution function of captured electrons, $f_s$, is defined
by the balance equation for them,
\begin{equation}\label{bal}
\sum_{\mathbf{k}}W_{\mathbf{k}s}f_s(1-f_{\mathbf{k}})=\sum_{\mathbf{k}}W_{s\mathbf{k}}f_{\mathbf{k}}(1-f_s),
\end{equation}
where
\begin{equation}\label{Wkss}
W_{s\mathbf{k}}=W_{\mathbf{k}s}=\frac{\hbar^2\Gamma_s}{Sm_e[(\varepsilon_k-\varepsilon_{s})^2+(\Gamma_s/2)^2]}
\end{equation}
is the probability of capture of a conduction electron with the
wave vector $\mathbf{k}$ and the energy
$\varepsilon_k=\hbar^2k^2/2m_e$ by a scatterer, which is derived
in Appendix B. Physically, the balance equation~(\ref{bal}) means
that the probability of electron transition from the bound state
to the continuum of free conduction electrons is equal to the
probability of the inverse process for any stationary distribution
of electrons. Substituting Eq.~(\ref{Wkss}) into Eq.~(\ref{bal})
and keeping in mind Eqs.~(\ref{sumk}) and (\ref{Ge}), the
distribution function of conduction electrons captured by
scatterers reads
\begin{equation}\label{fs}
f_s=\sum_{\mathbf{k}}W_{s\mathbf{k}}f_\mathbf{k}.
\end{equation}

Since an oscillating field driving 2D electrons is assumed to
satisfy the high-frequency condition (\ref{taue}), one can neglect
the collisional absorption of the field by conduction electrons.
Therefore, in the absence of the stationary field, $\mathbf{E}=0$,
the electronic system is in the thermodynamic equilibrium and the
distribution function is $f_{\mathbf{k}}=f^{(0)}(\varepsilon_k)$,
where
$$
f^{(0)}(\varepsilon_k)=\frac{1}{\exp[(\varepsilon_k-\varepsilon_F)/T]+1}
$$
is the Fermi-Dirac distribution function, $\varepsilon_F$ is the
Fermi energy, and $T$ is the temperature. Substituting the
distribution function $f_{\mathbf{k}}=f^{(0)}(\varepsilon_k)$ and
the probability (\ref{Wkss}) into Eq.~(\ref{fs}), we obtain the
equilibrium distribution function of conduction electrons captured
by the scatterers,
\begin{equation}\label{fss}
f^{(0)}_s=\frac{1}{\pi}\int_0^\infty\frac{(\Gamma_s/2)
f^{(0)}(\varepsilon_k)d\varepsilon_{k}}{(\varepsilon_k-\varepsilon_{s})^2+(\Gamma_s/2)^2}.
\end{equation}
Certainly, in the limiting case of stationary bound electron state
($\Gamma_s\rightarrow 0$), the distribution function (\ref{fss})
turns into the Fermi-Dirac function,
$f^{(0)}_s=f^{(0)}(\varepsilon_s)$. Substituting the equilibrium
distribution functions $f_{\mathbf{k}}=f^{(0)}(\varepsilon_k)$ and
$f_s=f^{(0)}_s$ into Eq.~(\ref{Be2}), we arrive at the equation
defining the Fermi energy $\varepsilon_F$,
\begin{equation}\label{EF}
n_sf^{(0)}_s+n_e=n_0,
\end{equation}
where $n_s=N_s/S$ is the density of scatterers, $n_0=N_e/S$ is the
density of conduction electrons in the absence of irradiation, and
\begin{equation}\label{ne}
n_e=\frac{2}{S}\sum_{\mathbf{k}}f^{(0)}(\varepsilon_k)=-\frac{m_eT}{\pi\hbar^2}\ln\left[\frac{1}{1+e^{\varepsilon_F/T}}\right]
\end{equation}
is the density of conduction electrons in the irradiated 2D
system.

Assuming the thermodynamic equilibrium to be weakly perturbed by
the stationary field $\mathbf{E}$, the sought distribution
functions can be written as
$f_{\mathbf{k}}=f^{(0)}(\varepsilon_k)+\Delta f_{\mathbf{k}}$ and
$f_s=f^{(0)}_s+\Delta f_s$, where $\Delta f_{\mathbf{k}}$ and
$\Delta f_s$ are small nonequilibrium additions arisen from the
field $\mathbf{E}$. To find these additions, let us assume that
the temperature is around zero in order to neglect the phonon
scattering of conduction electrons. Then the scattering
probability per unit time, $w_{\mathbf{k}\mathbf{k}^\prime}$, can
be written in the explicit form as
\begin{equation}\label{wkK}
w_{\mathbf{k}\mathbf{k}^\prime}=(1-f_s)w^{(1)}_{\mathbf{k}\mathbf{k}^\prime}+f_sw^{(2)}_{\mathbf{k}\mathbf{k}^\prime},
\end{equation}
where the first term,
\begin{eqnarray}\label{wkk1}
w^{(1)}_{\mathbf{k}\mathbf{k}^\prime}&=&\frac{2\pi
n_s}{S\hbar}\Bigg|\gamma_{\mathbf{k}\mathbf{k}^\prime}U_{\mathbf{k}\mathbf{k}^\prime}\nonumber\\
&+&\frac{\hbar^2\Gamma_s}{m_e(\varepsilon_k-\varepsilon_{s}+i\Gamma_s/2)}
\Bigg|^2\delta(\varepsilon_{{k}^\prime}-\varepsilon_{{k}}),
\end{eqnarray}
describes the scattering of conduction electrons by empty
scatterers, and the second term,
\begin{eqnarray}\label{wkk2}
w^{(2)}_{\mathbf{k}\mathbf{k}^\prime}&=&\frac{2\pi
n_s}{S\hbar}\Bigg|\gamma_{\mathbf{k}\mathbf{k}^\prime}U_{\mathbf{k}\mathbf{k}^\prime}+u_{\mathbf{k}\mathbf{k}^\prime}\nonumber\\
&-&\frac{\hbar^2\Gamma_s}{m_e(\varepsilon_k-\varepsilon_{s}+i\Gamma_s/2)}
\Bigg|^2\delta(\varepsilon_{{k}^\prime}-\varepsilon_{{k}}),
\end{eqnarray}
describes the scattering of conduction electrons by scatterers
containing captured electrons. Here
$U_{\mathbf{k}\mathbf{k}^\prime}$ is the matrix element of the
Born scattering by the initial potential $U(\mathbf{r})$,
$u_{\mathbf{k}\mathbf{k}^\prime}$ is the matrix element of the
Born scattering by the potential addition $u(\mathbf{r})$ produced
by a captured electron, and
$\gamma_{\mathbf{k}\mathbf{k}^\prime}=J_0[2kr_0\sin(\theta/2)]$
[see Appendix B for details of derivation of the probabilities
(\ref{wkk1}) and (\ref{wkk2})].

The first terms under modulus in Eqs.~(\ref{wkk1})--(\ref{wkk2})
describes the conventional potential scattering, whereas the last
therm corresponds to the electron scattering through the
quasi-discrete energy level $\varepsilon_s$ and is physically
identical to the Breit-Wigner equation for the resonant
scattering~\cite{Landau_3}. The probabilities
(\ref{wkk1})--(\ref{wkk2}) take into account the quantum
interference of these two scattering processes, which can manifest
itself in electron transport as features of the Fano
kind~\cite{Fano_1961}. However, the interference term is small
under the condition of small broadening, $\Gamma_s$, and,
therefore, can be neglected at a first approximation. It should be
noted also that the second terms under modulus in
Eqs.~(\ref{wkk1})--(\ref{wkk2}) are of different signs.
Physically, this originates from different intermediate states
involved in the resonant scattering described by Eqs.~(\ref{wkk1})
and (\ref{wkk2}): In the case of an empty scatterer the
intermediate state corresponds to the scatterer containing a
captured electron, whereas the intermediate state of a scatterer
containing a captured electron corresponds to the empty scatterer.

Since the probabilities (\ref{wkk1}) and (\ref{wkk2}) describe the
elastic scattering of conduction electrons with the isotropic
energy spectrum $\varepsilon_k=\hbar^2k^2/2m_e$ and the scattering
potentials $U(\mathbf{r})$ and $u(\mathbf{r})$ are assumed to be
axially symmetric in the 2D plane, the total probability
(\ref{wkK}) can be easily rewritten as a function of the angle
$\theta=\widehat{\mathbf{k},\mathbf{k}^\prime}$ and the modulus of
electron wave vector $k$. Therefore, it is convenient to write the
scattering matric elements as functions of these arguments,
$U_{\mathbf{k}\mathbf{k}^\prime}=U_k(\theta)$ and
$u_{\mathbf{k}\mathbf{k}^\prime}=u_k(\theta)$. Then the relaxation
time approximation is applicable to solve the Boltzmann kinetic
equation (\ref{Be1}) and the sought nonequilibrium distribution
functions read
\begin{eqnarray}\label{sol}
\Delta f_\mathbf{k}&=&\left[-\frac{\partial
f^{(0)}(\varepsilon_k)}{\partial\varepsilon_k}\right]e\tau_{k}\mathbf{v}_\mathbf{k}\mathbf{E},\label{sol1}\\
\Delta f_s&=&0,\label{sol2}
\end{eqnarray}
where
$\mathbf{v}_\mathbf{k}=\partial_{\mathbf{k}}\varepsilon_{\mathbf{k}}/\hbar$
is the velocity of conduction electron with the wave vector
$\mathbf{k}$, the transport relaxation time, $\tau_{k}$, is
defined by the expression
\begin{equation}\label{tau}
\frac{1}{\tau_{k}}=\int_{-\pi}^\pi(1-\cos\theta)w_k(\theta)d\theta,
\end{equation}
where the effective scattering probability per unit time,
$w_k(\theta)$, reads
\begin{widetext}
\begin{eqnarray}\label{wT}
w_k(\theta)&=&\frac{
n_sm_e}{2\pi\hbar^3}\left[(1-f_s)\Bigg|\left.J_0[2kr_0\sin(\theta/2)]U_{k}(\theta)+\frac{\hbar^2\Gamma_s}{
m_e(\varepsilon_k-\varepsilon_{s}+i\Gamma_s/2)}\Bigg|^2\right.\right.
\nonumber\\
&+&\left.f_s\Bigg|J_0[2kr_0\sin(\theta/2)U_k(\theta)+u_k(\theta)-\frac{\hbar^2\Gamma_s}{
m_e(\varepsilon_k-\varepsilon_{s}+i\Gamma_s/2)}\,\Bigg|^2\right].
\end{eqnarray}
\end{widetext}
Certainly, the distribution functions (\ref{sol1})--(\ref{sol2})
satisfy the Boltzmann kinetic equation (\ref{Be1}) and can be
easily verified by direct substitution into this. Substituting
Eqs.~(\ref{tau})--(\ref{wT}) into Eq.~(\ref{sol}) and summating it
over all electronic states, we arrive at the conductivity of the
2D system
\begin{equation}\label{sigma}
\sigma=\frac{e^2m_e}{2\pi\hbar^2}\int_0^\infty\left[-\frac{\partial
f^{(0)}(\varepsilon_k)}{\partial\varepsilon_k}\right]v^2_k\tau_k\,d\varepsilon_k.
\end{equation}

\section{Results and discussion}
For definiteness, let us restrict the analysis of electron
transport by the case of zero temperature, $T=0$, when the
conductivity (\ref{sigma}) reads
\begin{equation}\label{sigma0}
\sigma=\frac{e^2n_e\tau_F}{m_e},
\end{equation}
where $\tau_F$ is the relaxation time (\ref{tau}) for
$\varepsilon_k=\varepsilon_F$. Taking into account
Eqs.~(\ref{fss})--(\ref{EF}) and (\ref{Ge}), the density of
conduction electrons, $n_e$, is defined by the equation
\begin{equation}\label{neR}
n_e=n_0-\frac{n_s}{2}-\frac{n_s}{\pi}\tan^{-1}\left(\frac{2[\varepsilon_F-\varepsilon_s]}{\Gamma_s}\right),
\end{equation}
where $\varepsilon_F=\pi\hbar^2 n_e/m_e$ is the Fermi energy. In
what follows, we will be to consider the physically relevant case
of small density of scatterers, $n_s\ll n_0$. Then the solution of
Eq.~(\ref{neR}) reads
\begin{equation}\label{ne1}
n_e=
n_0-\frac{n_s}{2}-\frac{n_s}{\pi}\tan^{-1}\left(\frac{2[\varepsilon_{F0}-\varepsilon_s]}{\Gamma_s}\right),
\end{equation}
where $\varepsilon_{F0}=\pi\hbar^2 n_0/m_e$ is the Fermi energy in
the absence of irradiation. Substituting Eq.~(\ref{tau}) into
Eq.~(\ref{sigma0}) and keeping in mind that the broadening
$\Gamma_s$ is assumed to be very small, the resistivity of 2D
system, $\rho=1/\sigma$, can be written as a sum,
\begin{equation}\label{tau1}
\rho={\rho_{\mathrm{2D}}}+{\rho_{\mathrm{Q}}},
\end{equation}
where the first term describes the resistivity of 2D system arisen
from the usual potential scattering of conduction electrons,
whereas the second term,
\begin{equation}\label{tau2}
{\rho_{\mathrm{Q}}}=\frac{2n_s}{\pi
n_e}\left(\frac{h}{e^2}\right)\frac{(\Gamma_s/2)^2}{(\varepsilon_F-\varepsilon_{s})^2+(\Gamma_s/2)^2},
\end{equation}
is of purely quantum nature and describes the scattering of
conduction electrons trough the light-induced quasi-stationary
bound states (the Breit-Wigner resonant scattering), where $h/e^2$
is the resistivity quantum. It follows from (\ref{tau2}) that the
quantum resistivity depends resonantly on the Fermi
energy $\varepsilon_F$ with the resonance at
$\varepsilon_F=\varepsilon_s$ and the resonant amplitude is
$\bar{\rho}_{\mathrm{Q}}=(2n_s/\pi n_e)(h/e^2)$.

In the present analysis of electron transport, we assumed the most
general types of the scattering potential $U(\mathbf{r})$ and 2D
electron system. To proceed, we have to make some approximations.
Let us restrict the following analysis  by the case of short-range
scatterers which are conventionally modeled in 2D systems by the
delta-potential (see, e.g., Ref.~\onlinecite{Ando_1982}). This
corresponds, particularly, to a semiconductor quantum well doped
by neutral atoms. Then we can apply Eqs.~(\ref{ee0E})--(\ref{G2E})
to describe the energy of the quasi-stationary bound electron
state, $\varepsilon_s$, and its broadening, $\Gamma_s$. Next, let
us consider the 2D electron system in GaAs-based quantum well,
where conduction electrons fill only the ground subband (the
electron density is $n_0=5\cdot10^{11}\,\mathrm{cm}^{-2}$) and the
electron effective mass is $m_e\approx0.067\,m_0$ ($m_0$ is the
electron mass in vacuum). In such modern quantum wells, the
electron mobility is
$\mu=|e|\tau_e/m_e\sim10^6-10^7\,\mathrm{cm}^2/\mathrm{V}\cdot\mathrm{s}$.
Therefore, the condition (\ref{taue}) can be satisfied near the
high-frequency boarder of the microwave range. For instance, we
have $\omega\tau_e\sim10$ for the field frequencies around
$\nu=\omega/2\pi=100\,\mathrm{GHz}$. Therefore, the photon energy
of the dressing field can be chosen as
$\hbar\omega=1\,\mathrm{meV}$.
\begin{figure}[!ht]
\includegraphics[width=1.0\columnwidth]{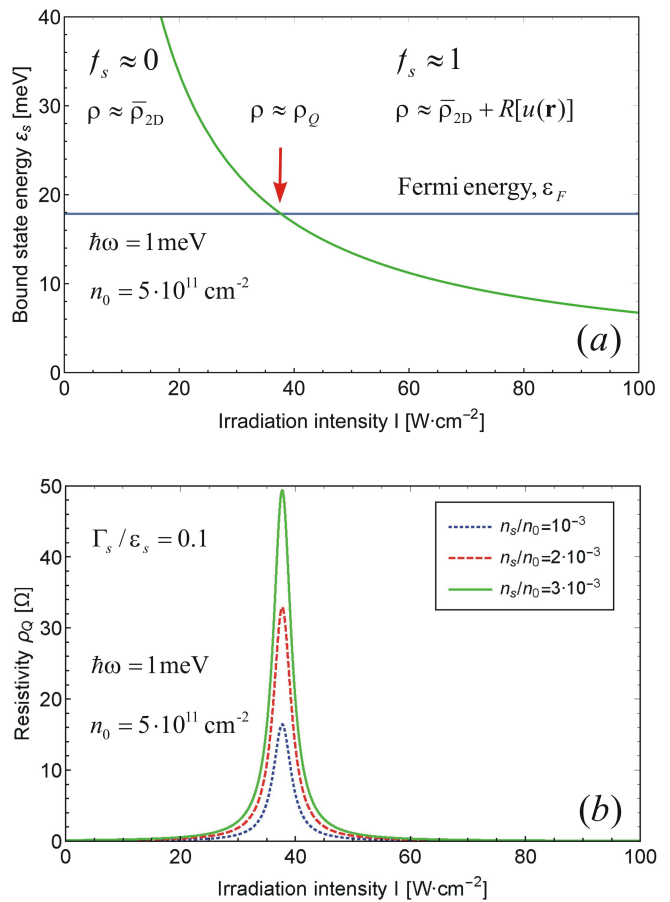}
\caption{Electronic characteristics of a GaAs-based quantum well
filled by conduction electrons with the density
$n_0=5\cdot10^{11}\,\mathrm{cm}^{-2}$ and irradiated by a
circularly polarized electromagnetic wave with the intensity $I$
and the photon energy $\hbar\omega=1\,\mathrm{meV}$: (a)
Dependence of the bound state energy, $\varepsilon_s$, on the
irradiation intensity, $I$, where the horizontal blue line marks
the Fermi energy, $\varepsilon_F$, and the red vertical arrow
marks the resonant point, $\varepsilon_s=\varepsilon_F$; (b)
Dependence of the quantum resistivity, $\rho_Q$, on the
irradiation intensity, $I$, for the energy broadening
$\Gamma_s=0.1\,\varepsilon_s$ and different scatterer densities,
$n_s$.}\label{Fig.2}
\end{figure}

The dependence of the bound state energy, $\varepsilon_s$, on the
irradiation intensity, $I=cE^2_0/4\pi$, is plotted in Fig.~2a,
where the vertical arrow marks the resonant point
($\varepsilon_s=\varepsilon_F$). Far from the resonant point
($|\varepsilon_s-\varepsilon_F|\gg\Gamma_s$), the resonant term
(\ref{tau2}) can be neglected and the total resistivity
(\ref{tau1}) is $\rho\approx{\rho_{\mathrm{2D}}}$. On the left of
the resonant point, the bound states are empty ($f_s\approx0$) and
this resistivity reads
${\rho_{\mathrm{2D}}}=\bar{\rho}_{\mathrm{2D}}$, where
\begin{equation}\label{rho0}
\bar{\rho}_{\mathrm{2D}}=\left(\frac{{\rho_0}}{2\pi}\right){\int_{-\pi}^\pi(1-\cos\theta)J^2_0[2k_Fr_0\sin({\theta}/{2})]d\theta},
\end{equation}
$\rho_0$ is the resistivity of the considered 2D system in the
absence of irradiation, and $k_F=\sqrt{2\pi n_e}$ is the Fermi
wave vector of conduction electrons. It should be noted that
Eq.~(\ref{rho0}) exactly coincides with the equation derived and
analyzed in Ref.~\onlinecite{Morina_2015} beyond the effect of
light-induced quasi-stationary electron states. On the right of
the resonant point ($f_s\approx1$), the bound states are filled by
captured electrons and the resistivity is
${\rho_{\mathrm{2D}}}=\bar{\rho}_{\mathrm{2D}}+R[u(\mathbf{r})]$,
where $R[u(\mathbf{r})]$ is the functional depending on the
scattering potential produced by a captured electron,
$u(\mathbf{r})$. Physically, the addition $R[u(\mathbf{r})]$ to
the resistivity (\ref{rho0}) arises from the fact that the
scatterers filled with captured electrons scatter conduction
electrons more effective than the empty scatterers. It should be
noted that the capture of conduction electrons by the
light-induced bound states leads also to the decreasing of density
of conduction electrons, $n_e<n_0$ [see
Eqs.~(\ref{neR})--(\ref{ne1})]. However, we have $n_e\approx n_0$
under the condition $n_s\ll n_e$ and, therefore, the contribution
of decreasing density of conduction electrons to increasing
resistivity is very small. Near the resonant point, we have to
take into account the resistivity term (\ref{tau2}). For typical
parameters of the modern GaAs-based quantum well (the density of
conduction electrons is
$n_0\approx5\cdot10^{11}\,\mathrm{cm}^{-2}$ and the electron
mobility is $\mu\approx10^{7}\,\mathrm{cm}^2/V\cdot\mathrm{s}$),
the resistivity of the system in the absence of irradiation is
$\rho_0\approx1\,\mathrm{\Omega}$, whereas the resistivity quantum
is $h/e^2\approx26\,\mathrm{k\Omega}$. Since $h/e^2\gg\rho_0$, the
resonant term (\ref{tau2}) is dominant near the resonant point in
the broad range of scatterer density, $n_s$. Therefore, we have
$\rho\approx\rho_Q$ there. The dependence of the quantum
resistivity (\ref{tau2}) on the irradiation intensity, $I$, is
plotted in Fig.~2b for different scatterer densities, $n_s$. It
should be noted that the resonance amplitude of the resistivity
(\ref{tau2}) is still large enough,
$\bar{\rho}_Q\sim10^{-1}\,\mathrm{\Omega}$, even if the scatterer
density is very small, $n_s\sim10^{-5}n_e$. Therefore, the quantum
correction to the resistivity (\ref{tau2}) can be detected in the
state-of-the-art measurements in the broad range of scatterer
densities.

In the present study, we considered the bound electron states
induced in 2D systems by a monochromatic electromagnetic wave. It
should be noted that such states take place in 1D systems as
well~\cite{Kibis_2019}. As to 3D systems, the bound states
disappear. Physically, this follows from the fact that alternating
electric field of the wave can localize an electron only along the
direction of its oscillations~\cite{Kibis_2019}. Since there are
no field oscillations along the direction of propagation of the
electromagnetic wave, the discussed mechanism of the light-induced
electron localization does not work in 3D systems.

\section{Conclusion}
Irradiation of a 2D electron system by a circularly polarized
off-resonant electromagnetic wave induces the quasi-stationary
electron states confined at repulsive scatterers and immersed into
the continuum of states of conduction electrons. These
quasi-stationary bound electron states result in the corrections
to conductivity of the irradiated 2D system through the two main
mechanisms: The capture of conduction electrons by the bound
states and the scattering of conduction electrons through these
states. As a consequence, the corrections to resistivity of two
kinds appear. The first of them is the non-resonant addition to
the resistivity arisen from the increasing of the scattering of
conduction electrons by the scatterers containing captured
electrons. The second is the resonant addition to resistivity of
purely quantum nature, which arises from the resonant Breit-Wigner
scattering of conduction electrons through the quasi-stationary
bound states (the resonant peak of the resistivity takes place
when the Fermi energy of conduction electrons coincides with the
quasi-discrete energy level of the bound state). Within the model
of short-range scatterers described by the repulsive
delta-potentials, the resistivity of GaAs-based quantum well is
studied in the broad ranges of irradiation intensity and scatterer
density.

\begin{acknowledgments}
The reported study was funded by the Russian Science Foundation
(project 20-12-00001).
\end{acknowledgments}

\appendix

\section{Energy and broadening of quasi-stationary electron states}

Let us consider the Schr\"odinger problem with the Hamiltonian
(\ref{H1}),
\begin{equation}\label{H11}
\hat{\cal
H}_0=-\frac{\hbar^2}{2m_er}\left[\frac{\partial}{\partial
r}\left(r\frac{\partial}{\partial
r}\right)+\frac{1}{r}\frac{\partial^2}{\partial\varphi^2}\right]+\frac{u_0\,\delta({r}-{r}_0)}{2\pi
r_0},
\end{equation}
where $\varphi$ is the azimuth angle in the 2D plane. The
eigenfunction of the Hamiltonian, $\psi$, which corresponds to an
electron confined in the area $0<r<r_0$, must be finite at $r=0$
and satisfy the condition $\psi|_{r\rightarrow\infty}\propto
e^{ikr}$, where $k=\sqrt{\,2m_e\varepsilon}/\hbar$ is the electron
wave vector and $\varepsilon$ is the electron energy. Therefore,
the sought wave function can be written with using the Bessel
functions as
\begin{equation}\label{Psi}
\psi=e^{im\varphi}\left\{
\begin{array}{rl}
AJ_m(kr), &0<r<r_0\\
CH_m(kr), &r>r_0
\end{array}\right.,
\end{equation}
where $m=0,\pm1,\pm2,...$ is the angular electron momentum,
$J_m(z)$ is the Bessel function of the first kind,
$H_m(z)=J_m(z)+iN_m(z)$ is the Bessel function of the third kind
(the Hankel function of the first kind), $N_m(z)$ is the Bessel
function of the second kind (the Neumann function), and $A$, $C$
are the constants. Substituting the wave function (\ref{Psi}) into
the Schr\"odinger equation with the Hamiltonian (\ref{H11}) and
integrating it over $r$ near ${r}= {r}_0$, we arrive at the
continuity condition for electron current density at the
ring-shape delta-potential barrier,
\begin{equation}\label{cc1}
CH^\prime_m(z)-A\left[J^\prime_m(z)+\frac{m_eu_0}{\pi\hbar^2kr_0}J_m(z)\right]=0,
\end{equation}
where $z=kr_0$. As to the continuity condition for the electron
wave function (\ref{Psi}) at the barrier, it reads
\begin{equation}\label{cc2}
CH_m(z)-AJ_m(z)=0.
\end{equation}
The two homogeneous algebraic equations (\ref{cc1}) and
(\ref{cc2}) define the constants $A$ and $C$, whereas the secular
equation arisen from them,
\begin{eqnarray}\label{sec}
H^\prime_m(z)J_m(z)-H_m(z)\left[J^\prime_m(z)+\frac{m_eu_0}{\pi\hbar^2kr_0}J_m(z)\right]=0,\nonumber\\
\end{eqnarray}
defines the total electron energy, $\varepsilon$. To simplify
Eq.~(\ref{sec}), let us apply the known equalities,
$J_{\nu}(z)N_{\nu-1}(z)-J_{\nu+1}(z)N_{\nu}(z)=-2/\pi z$ and
$Z^\prime_\nu(z)=[Z_{\nu-1}(z)-Z_{\nu+1}(z)]/2$, where
$Z_{\nu}(z)$ is any Bessel function (see, e.g.,
Ref.~\onlinecite{Gradstein_book}). Then Eq.~(\ref{sec}) can be
rewritten in the compact form as
\begin{equation}\label{sec0}
H_m(z)J_m(z)=i{\alpha},
\end{equation}
where $z=kr_0$ and $\alpha=2\hbar^2/m_eu_0$.

Although the secular equation (\ref{sec0}) can be easily solved
numerically, there is the important particular case of strong
repulsive potential, $\alpha\ll1$, when solution of this equation
can be found analytically. Namely, let us seek roots of
Eq.~(\ref{sec0}) as a power series,
$z_{nm}=\bar{z}_{nm}+\sum_{\,l=1}^{\,\infty}
z^{(l)}_{nm}\alpha^l$, where the integer $n$ numerates the roots
and $z^{(l)}_{nm}$ are the expansion coefficients. Substituting
$z=z_{nm}$ into Eq.~(\ref{sec0}) and expanding the Bessel
functions there into the Taylor series near $\alpha=0$, we arrive
at the system of algebraic recurrence equations for the expansion
coefficients. The solving of the system results in the sought
energy spectrum of an electron confined inside the ring-shape
delta-barrier, $\varepsilon=\varepsilon_{nm}-i\Gamma_{nm}/2$, and
the corresponding wave functions (\ref{Psi}), where the energy of
quasi-discrete electron level is
\begin{equation}\label{eee}
\varepsilon_{nm}=\frac{\hbar^2\bar{z}^2_{nm}}{2m_er^2_0}+{\cal
O}\left(\alpha\right),
\end{equation}
the broadening of the energy level is
\begin{equation}\label{G2}
{\Gamma}_{nm}=\frac{4\varepsilon_{nm}\alpha^2}{N^3_m(\bar{z}_{nm})[J_{m+1}(\bar{z}_{nm})-J_{m-1}(\bar{z}_{nm})]}+{\cal
O}\left(\alpha^3\right),
\end{equation}
the wave functions (\ref{Psi}) read
\begin{eqnarray}\label{FlmA}
\psi_{nm}&=&\frac{e^{im\varphi}}{\sqrt{\pi}r_0J_{m+1}(\bar{z}_{nm})}\left\{
\begin{array}{rl}
J_m\left(\frac{\bar{z}_{nm}r}{r_0}\right), &0<r\le r_0\\
0, &r\ge r_0
\end{array}\right.\nonumber\\
&+&{\cal O}\left(\alpha\right),
\end{eqnarray}
$\bar{z}_{nm}$ is the $n$th zero of the $m$th Bessel function of
the first kind (i.e. $J_m(\bar{z}_{nm})=0$), and $n=1,2,3,...$ is
the principal quantum number.

\section{Probabilities of electron transitions}

The interaction between the bound electron state,
$|s\rangle=\varphi_s(r)$, and the states of free conduction
electrons,
$|\mathbf{k}\rangle=\sqrt{1/S}e^{i\mathbf{k}\mathbf{r}}$, arises
from the electron tunneling through the potential barrier,
$|s\rangle\rightarrow|\mathbf{k}\rangle$ (see Fig.~1). Since the
tunneling is assumed to be weak, it can be described in the most
general form by the tunnel Hamiltonian
\begin{equation}\label{H}
\hat{\cal H}_T=|s\rangle\varepsilon_s\langle
s|+\sum_\mathbf{k}|\mathbf{k}\rangle\varepsilon_{k}\langle
\mathbf{k}|+\sum_\mathbf{k}\left[\,|\mathbf{k}\rangle|T_\mathbf{k}|\langle
s|+\mathrm{H.c.}\right],
\end{equation}
where $T_\mathbf{k}=\langle\mathbf{k}|\hat{\cal H}_T|s\rangle$ is
the tunnel matrix element of the Hamiltonian between the localized
and delocalized electron states. As to the summation over electron
states with wave vectors
$\mathbf{k}=(k_x,k_y)=(k\cos\theta,k\sin\theta)$ in the 2D system
of the area $S$, it is equal to the following integration:
\begin{equation}\label{sumk}
\sum_\mathbf{k}\rightarrow\frac{S}{(2\pi)^2}\int_{0}^\infty
kdk\int_{0}^{2\pi}d\theta.
\end{equation}
The wave function satisfying the Schr\"odinger equation with the
Hamiltonian (\ref{H}) can be written as
$|\psi\rangle=a_s(t)e^{-i\varepsilon_st/\hbar}|s\rangle+\sum_\mathbf{k}a_\mathbf{k}(t)e^{-i\varepsilon_{k}t/\hbar}|
{k}\rangle$. It should be noted that a scatterer is assumed to be
centered at $\mathbf{R}=0$, where $\mathbf{R}$ is the radius
vector of the scatterer position. In the most general case of
$\mathbf{R}\neq0$, the amplitudes $a_\mathbf{k}$  should be
multiplied with the phase factor $e^{i\mathbf{kR}}$. Substituting
this wave function into the Schr\"odinger equation,
$i\hbar\partial_t|\psi\rangle=\hat{\cal H}|\psi\rangle$, and
taking into account the approximate orthogonality of the basic
states, $\langle\mathbf{k}|s\rangle=0$, we arrive at the quantum
dynamics equations for the expansion coefficients $a_s(t)$ and
$a_\mathbf{k}(t)$,
\begin{eqnarray}
i\hbar\dot{a}_s(t)&=&e^{i(\varepsilon_s-\varepsilon_{k})t/\hbar)}\sum_\mathbf{k}
T^\ast_\mathbf{k}a_\mathbf{k}(t),\label{dyn1}\\
i\hbar\dot{a}_\mathbf{k}(t)&=&e^{i(\varepsilon_{k}-\varepsilon_s)t/\hbar)}T_\mathbf{k}a_s(t).\label{dyn2}
\end{eqnarray}
Let an electron be in the bound state at the time $t=0$, i.e.
$a_s(0)=1$ and $a_\mathbf{k}(0)=0$. Then the integration of
Eq.~(\ref{dyn2}) results in
\begin{equation}\label{b}
a_\mathbf{k}(t)=-\frac{iT_\mathbf{k}}{\hbar}\int_0^{\,t}e^{i(\varepsilon_{k}-\varepsilon_s)t^\prime/\hbar}a_s(t^\prime)dt^\prime.
\end{equation}
Since the considered system is axially symmetrical, the matrix
element $T_\mathbf{k}$ depends only on the electron energy,
$\varepsilon_{k}=\hbar^2 k^2/2m$ and, therefore, can be denoted as
$T_\mathbf{k}=T_{\varepsilon_{k}}$. Substituting Eq.~(\ref{b})
into Eq.~(\ref{dyn1}), we arrive at the expression
$$\dot{a}_s(t)=-\frac{1}{\hbar^2}\sum_\mathbf{k}|T_{\varepsilon_{k}}|^2\int_0^{\,t}e^{i(\varepsilon_{k}-\varepsilon_s)t^\prime/\hbar}a_s(t^\prime)dt^\prime,$$
which can be rewritten as
\begin{align}\label{a}
&\dot{a}_s(t)=\nonumber\\
&-\frac{Sm_e}{2\pi\hbar^4}\int_0^{\,\infty}
d\varepsilon_{k}\,\,|T_{\varepsilon_{k}}|^2\int_0^{\,t}e^{i(\varepsilon_s-\varepsilon_{k})(t-t^\prime)/\hbar}a_s(t^\prime)dt^\prime.
\end{align}
This is still an exact equation since we just replaced two
differential equations (\ref{dyn1})--(\ref{dyn2}) with one linear
differential-integral equation (\ref{a}). Next we make the
approximation. Namely, let us take into account that the tunneling
between the states $|s\rangle$ and $|\mathbf{k}\rangle$ is very
weak. Then the quantity $|T_{\varepsilon_{k}}|^2$ varies little
around $\varepsilon_{k}=\varepsilon_s$ for which the time integral
in Eq.~(\ref{a}) is not negligible. Physically, this means that
the condition (\ref{Ge}) is assumed to be satisfied and,
therefore, the energy of outgoing electron, $\varepsilon_{k}$, is
near the energy of bound state, $\varepsilon_s$.

As a consequence, one can make the replacement
$|T_{\varepsilon_{k}}|\rightarrow |T_{\varepsilon_s}|$ and replace
the lower limit in the $\varepsilon_{k}$ integration with
$-\infty$. Since this integration results in the delta function,
$$\int_{-\infty}^{\,\infty}e^{i(\varepsilon_s-\varepsilon_{k})(t-t^\prime)/\hbar}d\varepsilon_{k}=2\pi\hbar\,{\delta(t-t^\prime)},$$
Eq.~(\ref{a}) takes the form
\begin{equation}\label{a1}
\dot{a}_s(t)=-\frac{\Gamma_s}{2\hbar}\,a_s(t),
\end{equation}
where
\begin{equation}\label{Gamma}
\Gamma_s=\frac{Sm_e}{\hbar^2}|\,T_{\varepsilon_s}|^2
\end{equation}
is the energy broadening of the bound electron state. It follows
from Eq.~(\ref{Gamma}) that
\begin{equation}\label{a2}
a_s(t)=e^{-\Gamma_s t/2\hbar}.
\end{equation}

Now, we can calculate the probabilities $W_{\mathbf{k}s}$ and
$w_{\mathbf{k}\mathbf{k}^\prime}$ which appear in the Boltzmann
kinetic equation [see Eqs.~(\ref{sct}) and (\ref{fs})].
Substituting Eq.~(\ref{a2}) into Eq.~(\ref{b}), the amplitude of
emission of free electron with the wave vector $\mathbf{k}$ from
the bound state during the time $t$ reads
\begin{equation}\label{bk}
a_\mathbf{k}(t)=-T_{\varepsilon_{k}}\frac{e^{i(\varepsilon_{k}-\varepsilon_s)t/\hbar-\Gamma_s
t/2\hbar}-1}{\varepsilon_k-\varepsilon_{s}+i\Gamma_s/2}.
\end{equation}
Making the replacement $T_{\varepsilon_{k}}\rightarrow
T_{\varepsilon_s}$, the total amplitude of electron transition
from the state $|s\rangle$ to the state $|\mathbf{k}\rangle$ is
\begin{equation}\label{bkk}
a_\mathbf{k}(\infty)=\frac{T_{\varepsilon_s}}{\varepsilon_k-\varepsilon_{s}+i\Gamma_s/2}.
\end{equation}
Taking into account the reversibility of electron transitions and
using Eq.~(\ref{Gamma}), the total probability of capture of a
conduction electron with the wave vector $\mathbf{k}$ by a
scatterer,
$W_{s\mathbf{k}}=W_{\mathbf{k}s}=|a_{\mathbf{k}}(\infty)|^2$, can
be written as
\begin{equation}\label{Wks}
W_{s\mathbf{k}}=\frac{\hbar^2\Gamma_s}{Sm_e[(\varepsilon_k-\varepsilon_{s})^2+(\Gamma_s/2)^2]}.
\end{equation}

To describe the scattering of conduction electrons in the most
general form, let us add the scattering Hamiltonian,
\begin{equation}\label{HS}
\hat{\cal
H}_S=\sum_{\mathbf{k}^{\prime\prime}}\sum_{\mathbf{k}^{\prime\prime\prime}\neq\mathbf{k}^{\prime\prime}}|\mathbf{k}^{\prime\prime}\rangle
\langle\mathbf{k}^{\prime\prime}|U(\mathbf{r}-\mathbf{r}_0[t])|\mathbf{k}^{\prime\prime\prime}\rangle\langle\mathbf{k}^{\prime\prime\prime}|,
\end{equation}
to the tunnel Hamiltonian (\ref{H}), where
$U(\mathbf{r}-\mathbf{r}_0[t])$ is the oscillating potential
(\ref{Uosc}). Then the probability of electron scattering per unit
time between the states $|\mathbf{k}^{\prime}\rangle$ and
$|\mathbf{k}\rangle$, which takes into account both tunnel
transitions arisen from the Hamiltonian (\ref{H}) and the
potential scattering induced by the Hamiltonian (\ref{HS}), is
defined in the lowest order of the conventional perturbation
theory~\cite{Landau_3} as
\begin{eqnarray}\label{wkk}
w_{\mathbf{k}\mathbf{k}^\prime}&=&\frac{2\pi}{\hbar}\left|\sum_{j=1}^{N_s}\left[\frac{\langle\mathbf{k}|\hat{\cal
H}_T|s\rangle\langle s|\hat{\cal
H}_T|\mathbf{k}^\prime\rangle}{\varepsilon_k-\varepsilon_{s}+i\Gamma_s/2}+
\langle\mathbf{k}|U_0(\mathbf{r})|\mathbf{k}^{\prime}\rangle\right]\right.\nonumber\\
&\times&e^{i(\mathbf{k}^\prime-\mathbf{k})\mathbf{R}_j}
\Bigg|^2\delta(\varepsilon_{{k}^\prime}-\varepsilon_{{k}}),
\end{eqnarray}
where the summation index $j$ numerates different scatterers, and
$U_0(\mathbf{r})$ is the time-averaged oscillating potential
(\ref{U0}). The first term in the square brackets of
Eq.~(\ref{wkk}) corresponds to the tunnel electron scattering
through the quasi-discrete energy level $\varepsilon_s$. As to the
second term, it describes the elastic scattering of conduction
electrons by the oscillating potential (\ref{Uosc}) and reads
\begin{equation}\label{Bkk}
\langle\mathbf{k}|U_0(\mathbf{r})|\mathbf{k}^{\prime}\rangle=\frac{\gamma_{\mathbf{k}\mathbf{k}^\prime}}{S}
U_{\mathbf{k}\mathbf{k}^\prime},
\end{equation}
where $U_{\mathbf{k}\mathbf{k}^\prime}=\langle
e^{i\mathbf{k}\mathbf{r}}|U(\mathbf{r})|e^{i\mathbf{k}^\prime\mathbf{r}}\rangle$
is the matrix element of the Born scattering for the initial
potential $U(\mathbf{r})$, the Bessel-function factor,
$$\gamma_{\mathbf{k}\mathbf{k}^\prime}=J_0[2kr_0\sin(\theta/2)],$$
arises from oscillations of the scattering potential (\ref{Uosc}),
and $\theta=\widehat{\mathbf{k},\mathbf{k}^\prime}$ is the angle
between the wave vectors of incident and scattered electron waves.
Taking into account the random arrangement of scatterers and using
Eq.~(\ref{Gamma}), the summation over the index $j$ in
Eq.~(\ref{wkk}) results in
\begin{eqnarray}
w_{\mathbf{k}\mathbf{k}^\prime}&=&\frac{2\pi
n_s}{S\hbar}\Bigg|\gamma_{\mathbf{k}\mathbf{k}^\prime}U_{\mathbf{k}\mathbf{k}^\prime}\nonumber\\
&+&\frac{\hbar^2\Gamma_s}{m_e(\varepsilon_k-\varepsilon_{s}+i\Gamma_s/2)}
\Bigg|^2\delta(\varepsilon_{{k}^\prime}-\varepsilon_{{k}}).\label{wkkk}
\end{eqnarray}

It should be noted that the probabilities discussed above are
derived in the rotating reference frame (\ref{rf}), where the
scattering potential (\ref{Uosc}) oscillates. However, they have
the same form in the laboratory reference frame, where the kinetic
Boltzmann equation depending on these probabilities is written
(see Section III). Generally, such an invariance follows from the
fact that an unitary transformation of the Hamiltonian to the new
reference frame does not change matrix elements of physical
quantities derived with using the Hamiltonian. In the particular
case of the considered system, this invariance can be proved by a
direct calculation of the matrix element (\ref{Bkk}) in the
laboratory reference frame, where the oscillating potential
(\ref{Uosc}) turns into the stationary potential $U(\mathbf{r})$
but the plane electron waves $e^{i\mathbf{k}\mathbf{r}}$ and
$e^{i\mathbf{k}^\prime\mathbf{r}}$ should be replaced with the
corresponding Floquet functions [the eigenfunctions of the first
term of the Hamiltonian (\ref{HA})]. As expected, the calculation
results in the same matrix element (\ref{Bkk}). Details of the
calculation can be found in Appendix of
Ref.~\onlinecite{Morina_2015}, where the potential scattering of
conduction electrons in 2D systems driven by an oscillating field
was analyzed within the Floquet theory beyond the effect of
field-induced quasi-stationary electron states.

\end{document}